\documentclass{article}
\usepackage{amsmath}
        \newcounter{eqnletter}[equation]
\catcode`\@=11 \@addtoreset{equation}{section} \catcode`\@=12

\usepackage[dvips]{graphicx}
\usepackage{color}
\usepackage{epsfig}

\begin{document} 

 {\centerline	{\LARGE {   Incremental universality of Wigner
 random matrices
} }
 \vskip .3 cm

{\centerline {	Giovanni M. Cicuta$^*$
\footnote{giovanni.cicuta@gmail.com} and Mario Pernici$^{**}$
\footnote{mario.pernici@mi.infn.it} }} \vskip .3 cm {\centerline  { $^*$
Dept. of Physics, Univ. of Parma, Viale delle Scienze 7A, 43100 Parma,
Italy}} \vskip .3 cm
	{\centerline  {$^{**}$ Istituto Nazionale di Fisica Nucleare,
	Sezione di Milano ,}}
{	\centerline  {	Via Celoria 16, 20133 Milano, Italy}}

		\vskip 1cm 
		{\centerline{\textbf{Abstract}} } \vskip .4 cm

\normalsize

Properties of universality have essential relevance for the theory of random matrices usually called the Wigner ensemble.
The issue was analysed up to recent years with detailed and relevant results.
We present a slightly different view and compare the large-$n$ limit of connected correlators of distinct ensembles: universality has steps or degrees, precisely counted by the number of probability moments of the matrix entries, which match among distinct ensembles.

\section{Introduction}
 The developments of random matrix theory were parallel to discoveries of universal laws in the limit of infinite order of the matrices.\\
	In the decades '70, '80, '90 important developments were made in the theory of invariant ensembles. 
These are ensembles of random matrices where the probability law for the matrices of the ensemble  is invariant under the similarity transformation by matrices of a classical group.
Among the important discoveries, we mention the  topological expansion by G. 't Hooft \cite{ge} , the analytic limiting solution of a generic one-matrix ensemble \cite{bre1}, the description of two-dimensional quantum gravity as a randomly triangulated manifold. A few references may provide help  \cite{amb}, \cite {diF} to recover the impressive developments over a few decades. 
A surprising result by E. Brezin and A. Zee \cite{bre2} proved that connected correlators between two finitely separated eigenvalues, when suitably smoothed, exhibit a higher level of universality than the density of eigenvalues.
  This universality law is very different from the \textsl{short distance} universality most studied, where the distance between the pair of eigenvalues in the correlation function is
 comparable to the spacing between adjacent eigenvalues. In this case, it is expected that the correlation function may be controlled by the (universal) level repulsion then originating the \textsl{short distance} universality  law, which was  called, for a long time, the Wigner-Dyson-Gaudin-Mehta conjecture \cite{yau}.\\
The proof of universality of the \textsl{short distance} correlations of eigenvalues had  several different histories.  For the case of invariant ensembles, one may see reference  \cite{dei} and we shall not mention invariant ensembles any further, because this note is pertinent only to Wigner ensembles.
These are non-invariant ensembles of random matrices, where the matrix entries are independent identically distributed random variables. For a few decades, universality was intended to mean, that the eigenvalue spectrum of Wigner matrices reproduced the properties of the Gaussian ensembles, for a large class of non-Gaussian measures \cite{ben} of the matrix entries.  Indeed when the asymptotic eigenvalue correlator was first computed for a generic Wigner ensemble, it was found that the leading term of the correlator for a pair of eigenvalues is a sum of a part equal to the Gaussian case, and a new one proportional to the fourth cumulant of the matrix entries probability law. This was interpreted as a limitation of universality \cite{az}, \cite{kkp1}, \cite{kkp2}.\\
 It is very reminiscent of the four moments theorem \cite{tao} asserting that the fine statistics of eigenvalues in the bulk of the spectrum  of a Wigner random matrix are only sensitive to the first four moments of the entries. The theorem is considered an important step in a sequence of works by mathematicians to control the fluctuations of eigenvalues on the local scale \cite{flo}.\\

	This note presents a general picture of universality for ensembles of Wigner random matrices. It is common sense that if distinct ensembles share several matching moments of the probability density of the matrix entries, then in the $n \to \infty$ limit, spectral density and the connected Green functions share a degree of universality. The present note is devoted to make explicit and quantitative this relation.  A list of the new assertions is summarized in the Conclusions.\\
	The combinatorial derivations and the result of exact evaluation of some relevant expectations are in the Supporting information appendix.\\

\section{Incremental universality of the single-trace averages.}

To illustrate the incremental universality in the simplest possible
setting, we consider an ensemble of real symmetric matrices $A$, of order
$n$. We assume all the diagonal entries to be zero. The off diagonal
entries $A_{i,j}$, $i < j$,  are
 independent identically distributed random variables. Their probability
 distribution $p(x)$ is even and has finite moments
    $v_{2k}= \langle (A_{i,j})^{2k}\rangle =\int x^{2k} p(x)\, d x $ of any order.\\
	As it was shown by E. Wigner, it is convenient to evaluate
the expectations of the traces $\langle \texttt{tr} A^{2k}\rangle$ by using a
correspondence with  closed walks of $2 k$ steps on the complete graph with 
$n$ vertices. The walks are grouped into  classes of equivalent walks \cite{alice}, \cite{ci}  
and by computer assisted algorithms one may evaluate these expectations, exact
for every $n > 2k$ , for small values of $k$.
The results may be written in  the form
\begin{equation}
\langle \texttt{tr} A^{2k}\rangle = \sum_{j=1}^k F_j^{(2k)}(v_2,\cdots,v_{2j}, n)
\label{s.1}
\end{equation}
where $F_j^{(2k)}( v_2, v_4,..,v_{2j}, n)$ is a finite sum
of monomials in the moments, each monomial having total moment degree $2k$ and
containing $v_{2j}$,
and with coefficient equal to an integer times $N_s$, with $2 \le s \le k + 2 - j$.
The falling factorial $N_s \equiv n(n-1)..(n-s+1)$ counts the number of walks
on $s$ distinct sites visited by a walk. 
For instance for $k=5$ one has
\begin{eqnarray}
&&F_1^{(10)} = (42 N_6 + 236 N_5 + 145 N_4)v_2^5 \nonumber \\
&&F_2^{(10)} = (120 N_5 + 385 N_4) v_4 v_2^3 + (65 N_4 + 90 N_3)v_4^2 v_2
\nonumber \\
&&F_3^{(10)} = (45 N_4 + 50 N_3) v_6 v_2^2 + 20 N_3 v_6 v_4 \nonumber \\
&&F_4^{(10)} = 10 N_3 v_8 v_2;  \quad
F_5^{(10)} = N_2 v_{10} 
\nonumber 
\end{eqnarray} 
We use the proper rescaling of the random matrices,  
$B=\frac{1}{\sqrt{(n-1) v_2}} A$.
The moments of the distribution of the entries of matrix $B$
are the ($n$-scaled) standardized moments $\tilde{v}_{2j}:= \frac{1}{(n-1)^j}\frac{v_{2j}}{v_2^j}$.

Under the assumption that the set of  moments $\{v_{2j} \}$ of matrix $A$ do not depend on $n$, it is convenient
to introduce
\begin{equation}
	g_j^{(2k)}(v_2,\cdots,v_{2j}, n) = 
	\frac{n^{j-2} F_j^{(2k)}(v_2,\cdots,v_{2j}, n)}{(n-1)^k \,v_2^k}
\end{equation}
which is finite for $n \to \infty$. Eq. (\ref{s.1}) becomes
\begin{equation}
\langle \frac{1}{n} \texttt{tr} B^{2k}\rangle = 
	 \sum_{j=1}^k n^{1-j} g_j^{(2k)}(v_2,\cdots,v_{2j},n)
\end{equation}
One has
\begin{eqnarray}
 &&\lim_{n \to \infty} g_j^{(2k)} (v_2, \dots , v_{2j},n)= \left\{
 \begin{array}{cccc} \frac{1}{k+1} \left( \begin{array}{ccc} 2k \\
 k \end{array} \right) \quad , \quad \texttt{if}\quad j=1 \\
\left( \begin{array}{ccc} 2k \\ k-j \end{array} \right)
\frac{v_{2j}}{v_2^j } \quad , \quad \texttt{if} \quad j=2, 3,..,k
\end{array} \right.  \nonumber\\ \label{s.2} 
\end{eqnarray}
The case $j=1$ has been computed by Wigner; the case $j > 1$ follows from the Theorem 
in Appendix B.

Define the class $T^{(j-1)}_{v_2,\cdots, v_{2(j-1)}}$ of Wigner random matrix ensembles, 
with the moments of the entries $v_2,\cdots, v_{2(j-1)}$.
Let us now consider two ensembles in this class; indicate with 
$\{ v_{2m}^{(1)}\} $ the set of moments of the probability law of the matrix 
entries of the first ensemble and $\{v_{2m}^{(2)}\} $ , the set of moments 
of the second ensemble. Let us suppose that all moments 
$\{ v_{2m}^{(1)}\} =\{ v_{2m}^{(2)}\}$ for $m=1,2,\dots ,j-1$, 
but they are different for $m=j > 1$.  Then

\begin{eqnarray}
        && \langle \frac{1}{n} \texttt{tr} B^{2k}\rangle_1-
	\langle \frac{1}{n} \texttt{tr} B^{2k}\rangle_2 = \qquad \nonumber \\
 &&\sum_{i=j}^k n^{1-i} 
\big(g_i^{(2k)}(v_2^{(1)},\cdots,v_{2i}^{(1)},n) -
 g_i^{(2k)}(v_2^{(2)},\cdots,v_{2i}^{(2)},n)\big) = \nonumber \\
 &&\frac{1}{n^{j-1} } \left( \begin{array}{ccc} 2k \\
k-j \end{array} \right)
\left( \frac{v_{2j}^{(1)} - v_{2j}^{(2)}}{v_2^j}+O(\frac{1}{n}) \right) \quad , \quad k\geq j \qquad
\label{s.3} \end{eqnarray}
while this difference vanishes for $k < j$.

For the classes of Wigner ensembles which share a set of moments of the matrix entries,
one has $T^{(j)}_{v_2,\cdots, v_{2 j}} \subset T^{(j-1)}_{v_2,\cdots, v_{2(j-1)}}$.
Decreasing $j$ by $1$ the size of the class increases, while 
the shared $1/n$-expansion of single-trace averages has one term less.
Perhaps one may call  incremental universality  this phenomenon. \\

 For the pair of ensembles previously considered in $T^{(j-1)}_{v_2,\cdots, v_{2(j-1)}}$
the difference of the spectral functions
$\Delta \rho_n^{(j)}(y)=\rho_n^{(1)}(y)-\rho_n^{(2)}(y)$ satisfies

\begin{equation}
\int_{-2}^2 dy \Delta \rho_n^{(j)}(y) y^{2k} = 
\langle \frac{1}{n} \texttt{tr} B^{2k}\rangle_1-
        \langle \frac{1}{n} \texttt{tr} B^{2k}\rangle_2 =
	n^{1-j} \frac{\Delta v_{2j}}{v_2^j}\binom{2k}{k-j} + O(n^{-j})
\label{drho}
\end{equation}
for integer $k \ge 0$.

Define
\begin{equation}
	R_j(y) = \frac{1}{2\pi}\big(y U_{2j-1}(\frac{y}{2}) - 2 U_{2j-2}(\frac{y}{2})\big)
	\frac{1}{\sqrt{4-y^2}}
\label{Aj}
\end{equation}
where $U_i$ is a  Chebyshev polynomials of the II kind.

In Appendix B we prove that, for $j \ge 1$,
\begin{equation}
        \int_{-2}^2 dy R_j(y) y^{2k} = \binom{2k}{k-j}
\label{Rj}
\end{equation}

From Eqs. (\ref{drho}), (\ref{Rj}) we formally get, for $j \ge 2$,
\begin{equation}
\Delta \rho_n^{(j)}(y) = n^{1-j} \frac{\Delta v_{2j}}{v_2^j} R_j(y) + O(n^{-j})
\label{s.5}
\end{equation}

For $j=2$ Eq. (\ref{s.5}) agrees with Th. $1.1$ in \cite{em}.

In the next Section the expectation of the product of two or more
traces is shown to have a structure completely analogous to equations
(\ref{s.1}), (\ref{s.2}), then leading to the stepwise universality, here
described. The correlators, or connected expectations, have subtractions
which cancel leading terms and introduce negative contributions in
the results.\\

\section{$r$-trace connected correlators for $r \ge 2$}
In Appendix B we show that, at order $n^{2-r-j}$ in the $1/n$ expansion
of an $r$-trace connected correlator, the term with highest moment is 
$v_{2j} v_2^{k-j}$, where $2k$ is the total degree of the correlator; 
the coefficient of this term has been computed. From this it follows that
the difference between two connected correlators, belonging to Wigner ensembles
with moments $v_{2m}$ equal for $m < j$ and different for $m=j \ge r$ satisfies
\begin{eqnarray}
&&\lim_{n \to \infty} n^{r+j-2}
\Delta \langle \texttt{tr} \frac{B^{m_1}}{n} \cdots \texttt{tr} \frac{B^{m_r}}{n}\rangle_c = \nonumber \qquad \\
&& 2^{r-1} \frac{ \Delta(v_{2j})}{v_2^j} 
\sum_{j_1=1}^{\frac{m_1}{2}}\cdots \sum_{j_r=1}^{\frac{m_2}{2}}
\delta_{j_1+\cdots+j_r, j} \binom{m_1}{\frac{m_1}{2} - j_1}\cdots 
\binom{m_r}{\frac{m_r}{2} - j_r} \qquad
\label{DBr}
\end{eqnarray}
for all $m_i$ even, $0$ otherwise.

The difference between the spectral densities of two ensembles in 
$T^{(j-1)}_{v_2,\cdots, v_{2(j-1)}}$ is
\begin{equation}
\int_{-2}^2 dy_1 \cdots \int_{-2}^2 dy_r \Delta \rho_n^{(j)}(y_1,\cdots,y_r) 
	y_1^{2k_1}\cdots y_r^{2k_r} = 
\Delta \langle \texttt{tr} \frac{B^{2k_1}}{n} \cdots \texttt{tr} \frac{B^{2k_r}}{n}\rangle_c
\label{drhor}
\end{equation}

From Eqs. (\ref{Rj},\ref{DBr}, \ref{drhor}) we formally get
\begin{eqnarray}
	&&\Delta \rho_n^{(j)}(y_1,\cdots,y_r) = \qquad \qquad \nonumber\\
&&	2^{r-1} n^{2-r-j} \frac{\Delta(v_{2j})}{v_2^j} 
 \sum_{j_1=1}^{k_1}\cdots \sum_{j_r=1}^{k_r} \delta_{j_1+\cdots+j_r, j}
R_{j_1}(y_1)\cdots R_{j_r}(y_r) + O(n^{1-r-j}) \nonumber\\
\label{drhor2}
\end{eqnarray}

For $r=2$ and $j_1=j_2=1$ Eq. (\ref{drhor2}) agrees with Eq. (1.5) in \cite{az} 
(apart from
a factor $v_2=\sigma^2$ since in that reference $\rho_n$ is a function of
$\mu = \sigma y_1$ and $\nu = \sigma y_2$).\\

\section{Conclusions}
The present work has considered the most simple ensemble of Wigner random matrices: real symmetric random matrices with zero diagonal entries, and the independent identically distributed off-diagonal entries have symmetric probability density with (even) moments of every order \footnote{
  The contribution of the diagonal entries is considered in Appendix C and D,
  to compare with the literature on the leading order of the two-point connected correlation function for Wigner ensembles.
}.
It seems likely that the incremental universality, here described, also occurs in other important ensembles, like sparse matrices, band matrices with wide band, block matrices, complex hermitean matrices and it may be investigated in a similar way.\\

The focus of this note is the set of all connected correlators 
$\langle \frac{1}{n}\texttt{tr} \,B^{m_1} \dots \frac{1}{n} \texttt{tr} B^{m_r}\rangle_c$ 
where the normalized random matrix $B$ belongs to a Wigner ensemble 
and the identically independent distributed matrix entries of the random matrix 
$A$ have a probability density with $n$-independent moments
  $\{v_2, \cdots, v_{2j},\cdots\}$, corresponding to standardized moments
  $ \{1, \tilde{v}_4, \dots , \tilde{v}_{2j}, \dots \} $.

Let us summarize some previous knowledge pertinent to our subject, to elucidate the new contributions.\\

\begin{itemize}

\item
	It has been known for a long time that the leading order in  $1/n$ expansion of $\langle \frac{1}{n} \texttt{tr} \,B^{2k} \rangle=\frac{1}{k+1} \left( \begin{array}{ccc} 2k\\k \end{array} \right)$. At the next order of the expansion it depends on $\tilde{v}_4$ \cite{em}.\\
	The evaluation of the first term $\tilde{v}_{2j}$ in the $1/n$ expansion, Eq. (\ref{s.2}),
allows to evaluate the first term which is different, Eq. (\ref{s.3}),
for two distinct random matrix ensembles, which share a set of moments of the entries,  and the first different contribution for the spectral densities, Eqs. (\ref{drho})-(\ref{s.5}).

 \item  
The leading order $n^{-2}$, in the $1/n$ expansion,  of the connected two-point correlators had been computed and shown to depend on $\tilde{v}_4$ \cite{kkp2}, \cite{az}.
The leading order $n^{-4}$ of the connected three-point correlators depends on $\tilde{v}_4$ and $\tilde{v}_6$ \cite{az}.
In this note, for any $j \ge r$ the term of order $n^{2-j-r}$ and containing
$\tilde{v}_{2j}$,  in the $1/n$ expansion for the $r$-trace correlators is evaluated.
All other terms in the $1/n$ expansion, till order $n^{2-j-r}$ included, 
depend only on the moments $\tilde{v}_4,\cdots, \tilde{v}_{2(j-1)}$.
The class of  Wigner random matrix ensembles with this sequence of moments
has a degree of universality, which may perhaps be called incremental
universality.
The difference between a correlator for two elements of this class satisfies
  Eq. (\ref{DBr}), the difference between the spectral densities Eq. (\ref{drhor2}).

 \item
In considering the asymptotic behavior of the full set of connected correlators, in the large-$n$ limit, for all distinct ensembles, the Gaussian ensemble, with its standardized moments $\{1, 3,\dots, (2j-1)!!, \dots \}$ has no special role.   Of course the Gaussian ensembles has unique properties that allow analytic evaluations. Both these remarks also hold  for the invariant random matrix ensembles. 

 \end{itemize}

The Supporting information appendix describes the derivation of the statements in the paper and the comparison with works of authors who performed evaluations in part overlapping with ours. It also includes the exact evaluations of some expectations for any $n$. These are useful low-order checks of the combinatorial analysis at every order.\\

\newpage
\section{Supporting information Appendix, Introduction and Index}

In the appendices we describe a way to compute the correlators of the Wigner
random matrix model dealt with in this note.

The appendices are the following:

A)  Connected correlation functions and walks

B) Leading order of the $r$-trace connected correlator contributions containing $v_{2j}$

C) $1/n$ order of the one-point Green function

D) Leading order of the two-point connected correlation function

E) Exact connected correlators at low orders.
\\
In Appendix A we review the definition of the connected correlators and discuss
their computation in term of walks.

In Appendix B we enumerate the contributions to correlators in which the walks
are on tree graphs, and in which a single edge is run over more than twice.
 
In Appendices C and D we rederive
some results in the literature using the formalism introduced in the first
two appendices. 
While in the rest of the paper we will concentrate on the model described in the 
introduction, we mention in these two appendices random matrices having also diagonal 
terms, to compare with the literature.

Appendix E contains the evaluation of some exact correlators,
in which we identify the contributions computed in the previous appendices.

\section{Appendix A: Connected correlators and walks}
In this appendix we discuss how to express correlators
\begin{equation}
	\langle \texttt{tr} A^{k_1} \cdots \texttt{tr} A^{k_r}\rangle 
\end{equation}
and the corresponding connected correlators in terms of walks.
For this purpose we use  an algorithm similar to the
"label and substitute algorithm" in \cite{bau} to separate the sum of indices appearing in the
traces in sums of indices all different one from the other. 
The first subsection ends with a proof that the $r$-trace connected correlator is depressed
by a factor $n^{2-2r}$ with respect to the corresponding correlator.

\subsection{Review of connected correlators}
The relation between expectation functions and their connected parts may be written in the well known formalism of Green's functions of quantum field theory.
We define formal series
$$Z(x_1, x_2, \dots , x_k, \dots)=\int e^{ x_1 \texttt{tr} A+x_2 \texttt{tr}A^2+\dots +x_k \texttt{tr}A^k+\dots} \, \prod_{i <j} p(A_{i,j}) \, d A_{i,j}$$
where $p(x)$ is the probability density of each independent entry of the random matrix. Expectation of products of traces are
$$\langle \texttt{tr}A^{k_1} \texttt{tr}A^{k_2} \dots \texttt{tr}A^{k_r} \rangle=
\frac{\partial^r}{\partial x_{k_1} \partial x_{k_2} \dots \partial x_{k_r}} Z(x_1, x_2, \dots , x_k , \dots) \bigg|_{x_j=0, \forall j}$$

Connected correlators are generated by $ W(x_1, x_2, \dots , x_k , \dots) = \log Z(x_1, x_2, \dots , x_k , \dots) $
$$\langle \texttt{tr}A^{k_1} \texttt{tr}A^{k_2} \dots \texttt{tr}A^{k_r} \rangle _c=
\frac{\partial^r}{\partial x_{k_1} \partial x_{k_2} \dots \partial x_{k_r}} W(x_1, x_2, \dots , x_k , \dots) \bigg|_{x_j=0, \forall j}$$

Let us use the shortening
\begin{eqnarray}
Z_{1,..,r}&=&\frac{\partial^r}{\partial x_{k_1} \partial x_{k_2} \dots \partial x_{k_r}} Z(x_1, x_2, \dots , x_k , \dots) \, , \nonumber\\
W_{1,..,r}&=&\frac{\partial^r}{\partial x_{k_1} \partial x_{k_2} \dots \partial x_{k_r}} W(x_1, x_2, \dots , x_k , \dots) \, , \nonumber\\
&\texttt{then} & \nonumber\\
 W_1 = Z^{-1} Z_1  &,&  W_{1,2} = Z^{-1} Z_{1,2} - W_1 W_2 \qquad \nonumber\\
 W_{1,2,3} &=&      Z^{-1} Z_{1,2,3} - Z^{-2}(Z_{2,3} Z_1 + Z_{1,3} Z_2 + Z_{1,2} Z_3) +
        2 \, Z^{-3} Z_1 Z_2 Z_3 =  \qquad
\nonumber  \\
        &&      Z^{-1}Z_{1,2,3} - W_{1,2} W_3 - W_{1,3} W_2 - W_{2,3} W_1 - W_1 W_2 W_3 \qquad
        \label{w2c}
\end{eqnarray}

For instance
\begin{eqnarray}
\langle \texttt{tr} A^{k_1} \texttt{tr} A^{k_2} \rangle_c &=&
-\frac{1}{Z^2(x_{k_1},x_{k_2})}\frac{\partial Z(x_{k_1},x_{k_2})}{\partial x_{k_1}}\frac{\partial Z(x_{k_1},x_{k_2})}{\partial x_{k_2}} \bigg|_{x_j=0, \forall j}  \nonumber\\
&+& \frac{1}{Z(x_{k_1},x_{k_2})}\frac{\partial^2}{\partial x_{k_1}\partial x_{k_2} }
Z(x_{k_1},x_{k_2})\bigg|_{x_j=0, \forall j}\nonumber\\
&=&     \langle \texttt{tr} A^{k_1} \texttt{tr} A^{k_2}\rangle -\langle \texttt{tr} A^{k_1}\rangle \langle \texttt{tr} A^{k_2} \rangle
\nonumber
 \end{eqnarray}

 $Z_{1,..,r}\,|_{x_j=0, \forall j}$ are the $r$-point correlators,  $W_{1,..,r}\,|_{x_j=0, \forall j}$  are the $r$-point  connected correlators.
 
Correlators may be evaluated in terms of the contributions corresponding
to walks. 
In the case of the two-point correlators,
        the paths contributing to a correlator consists in two walks,
\begin{eqnarray}
\langle \texttt{tr} A^{k_1} \texttt{tr} A^{k_2} \rangle_c =
        \sum_{\gamma_1, \gamma_2} 
        \langle (\texttt{tr} A^{k_1})_{\gamma_1} (\texttt{tr} A^{k_2})_{\gamma_2} \rangle -
        \langle (\texttt{tr} A^{k_1})_{\gamma_1} \rangle \langle (\texttt{tr} A^{k_2})_{\gamma_2} \rangle
\label{w2ca}
\end{eqnarray}

where $(\texttt{tr} A^{k_i})_{\gamma_i}$ is the product of matrix elements along
the walk $\gamma_i$.
The measure for the average on Wigner matrices is the product of the measures on
the distinct edges of the walk, so that if $\gamma_1$ and $\gamma_2$ do not have
an edge in common (but they can have vertices in common),
$\langle (\texttt{tr} A^{k_1})_{\gamma_1} (\texttt{tr} A^{k_2})_{\gamma_2} \rangle = 
 \langle (\texttt{tr} A^{k_1})_{\gamma_1} \rangle \langle (\texttt{tr} A^{k_2})_{\gamma_2} \rangle$.
Therefore the sum over $\gamma_1$ and $\gamma_2$  in Eq. (\ref{w2ca})
can be restricted to the case in which $\gamma_1$ and $\gamma_2$  have at least
one edge in common.

The three-point connected correlator is given by
\begin{eqnarray}
 \langle \texttt{tr} A^{k_1} \texttt{tr} A^{k_2} \texttt{tr} A^{k_3} \rangle_c &=&
\langle \texttt{tr} A^{k_1} \texttt{tr} A^{k_2} \texttt{tr} A^{k_3} \rangle -
\langle \texttt{tr} A^{k_1} \rangle \langle \texttt{tr} A^{k_2} \texttt{tr} A^{k_3} \rangle - \nonumber \\
&& \langle \texttt{tr} A^{k_2} \rangle \langle \texttt{tr} A^{k_1} \texttt{tr} A^{k_3} \rangle -
\langle \texttt{tr} A^{k_3} \rangle \langle \texttt{tr} A^{k_1} \texttt{tr} A^{k_2} \rangle + \nonumber \\
&& 2\langle \texttt{tr} A^{k_1} \rangle \langle \texttt{tr} A^{k_2} \rangle \langle \texttt{tr} A^{k_3} \rangle
\end{eqnarray}

Equivalently one can replace $\texttt{tr} A^{k_i}$ with $ (\texttt{tr} A^{k_i})_{\gamma_i}$
in this formula and sum over $\gamma_1, \gamma_2, \gamma_3$, similarly to
Eq. (\ref{w2ca}).

Define for short
\begin{equation}
\phi_i = (\texttt{tr} A^{k_i})_{\gamma_i}
\end{equation}
the product of matrix elements along the walk $\gamma_i$.
\begin{eqnarray}
\langle \phi_1 \phi_2 \phi_3 \rangle_c &=&
\langle \phi_1 \phi_2 \phi_3 \rangle - 
\langle \phi_1 \phi_2 \rangle \langle  \phi_3 \rangle - \nonumber \\
&& \langle \phi_1 \phi_3 \rangle \langle  \phi_2 \rangle - 
\langle \phi_2 \phi_3 \rangle \langle  \phi_1 \rangle +
2 \langle \phi_1 \rangle \langle  \phi_2 \rangle \langle \phi_3 \rangle
\label{phi3}
\end{eqnarray}
From Eq. (\ref{phi3}) one sees that if $\gamma_3$ has no edge in common
with $\gamma_1$ or $\gamma_2$, then the contribution due to these walks vanishes.

For $r \ge 3$
\begin{equation}
W_{1,\cdots,r} = \frac{\partial}{\partial x_r}(Z^{-1} Z_{1,\cdots, r-1} - \sum \prod W_{\cdots})
\end{equation}
In general one gets
\begin{equation}
Z^{-1} Z_{1,\cdots, r} = W_{1,\cdots,r} + \sum \prod W_{\cdots}
\label{wpart}
\end{equation}
in which the second term is on the proper subsets of $\{1,\cdots, r\}$.
Since the set of all the correlators $\langle \phi_1 \cdots \phi_r \rangle$
is the sum of $\langle \phi_1 \cdots \phi_r \rangle_c$ and of all the non-edge
connected correlators, from the fact that $W_{1,2}|_{x=0}$ is edge-connected and 
Eq. (\ref{wpart})
one obtains by induction that $W_{1,\cdots,r}|_{x=0}$ is edge-connected. 

To find out the leading order of an $r$-connected correlator, one can take
$r$ Wigner trees and get an edge-connected path out of them. Representing a
Wigner tree as a point and the edge connecting it to another Wigner tree as a line,
one gets a connected tree graph with $r$ vertices and $r-1$ edges.
Therefore separating the Wigner trees such that they have no vertex in common
with the other Wigner trees, one gets $2(r-1)$ more vertices.
Since the latter have  finite contribution  for $n \to \infty$, the connected $r$-trace
correlators in a Wigner random matrix model correspond to contributions of order $O(n^{2-2r})$ for $n \to \infty$. This has been previously proven in \cite{az}, \cite{kkp1}, \cite{kkp2}, \cite{alice}, \cite{tao2}.

In the next two sections we will compute the contribution to a $r$-connected correlator
due to paths consisting of $r$ Wigner trees which have a single edge in common. 

\subsection{Enumeration of walks}

To evaluate the correlators we separate the indices in indices which are all
different, as in \cite{bau}. 
Label the indices all different from each another as $i_s$, with $s$ indicating
the order of first appearence of an index; let us call $s$ a reduced index.
To a matrix element $A_{i_r,i_s}$ corresponds an oriented step $(r,s)$ of a walk.
To produce all the walks corresponding to $\texttt{tr} A^k$, generate iteratively
the terms $(A^k w)_{i_0}$, where $w$ is a vector; at each step the last reduced index 
can be one of the previous reduced indices or a new reduced index, exceeding by 
$1$ the maximum previous reduced index.
At each iteration step substitute $w$ with $A w$.
Let us use the notation in which the product of two edges defines a sequence
of edges, the sum of two products of indices indicates two sequences of edges.
\begin{eqnarray}
&&w_0 \nonumber \\
&&e_{0,1} w_1 \nonumber \\
&& e_{0,1} (e_{1,0} w_0 + e_{1,2} w_2)\nonumber \\
&& e_{0,1} e_{1,0} (e_{0,1}w_1 + e_{0,2}w_2) + e_{0,1} e_{1,2} (e_{2,0}w_0 + e_{2,1}w_1 + e_{2,3}w_3)
\nonumber \\
&& e_{0,1} e_{1,0} e_{0,1}(e_{1,0}w_0 + e_{1,2}w_2) +
e_{0,1} e_{1,0} e_{0,2} ( e_{2,0}w_0 + e_{2,1}w_1 + e_{2,3}w_3) + \nonumber \\
&& e_{0,1} e_{1,2} e_{2,0}(e_{0,1}w_1 + e_{0,2}w_2 + e_{0,3}w_3) +
e_{0,1} e_{1,2} e_{2,1} (e_{1,0}w_0 + e_{1,2}w_2 + e_{1,3}w_3) + \nonumber \\
&& e_{0,1} e_{1,2} e_{2,3}(e_{3,0}w_0 + e_{3,1}w_1 + e_{3,2}w_2 + e_{3,4}w_4)
\label{Av0}
\end{eqnarray}

From the last sum, taking $w_s = \delta_{s,0}$ one obtains
\begin{eqnarray}
\Gamma^{(0)}_4 = \{e_{0,1} e_{1,0} e_{0,1}e_{1,0},\, e_{0,1} e_{1,0} e_{0,2} e_{2,0},\,
e_{0,1} e_{1,2} e_{2,1} e_{1,0},\, e_{0,1} e_{1,2} e_{2,3}e_{3,0} \} \qquad
\label{Av04}
\end{eqnarray}
which gives the walks $\Gamma^{(0)}_4$ corresponding to $\texttt{tr} A^4$.
Similarly one generate the walks $\Gamma^{(0)}_k$ corresponding to $\texttt{tr} A^k$.
Taking the average, each of the mappings of a given walk with $V$ vertices
give the same result, so that one gets a factor $N_V$.

One can write
\begin{equation}
\texttt{tr} A^{k} = 
	\sum_{\gamma \in M \Gamma^{(0)}_{k}} 
(\texttt{tr} A^{k})_{\gamma}
\label{akg}
\end{equation}
where $M$ is the isomorphic mapping $M$ of the reduced indices $0,\cdots,V-1$
to the class of all the injective mappings $s \to i_s$, with $i_s \in \{1,\cdots,n\}$

Let us now consider the product of two traces. After expanding the sum in
$\texttt{tr} A^{k_1}$ as described above, for each term 
$(\texttt{tr} A^{k_1})_{\gamma_1}$ 
with distinct reduced indices
$0,1,\cdots,s$, the first reduced index of the next trace can be one of the previous 
reduced indices, or the new reduced index $s+1$. Then proceed in a similar way 
iteratively as before; the last reduced index of the second trace is then set equal
to its first reduced index.

One has for instance
\begin{equation}
\texttt{tr} A^2 \texttt{tr} A^2 = 
        \sum_{(\gamma_1,\gamma_2) \in M \Gamma^{(0)}_{2,2}} (\texttt{tr} A^2)_{\gamma_1} (\texttt{tr} A^2)_{\gamma_2}
\label{a22c}
\end{equation}

where
\begin{eqnarray}
&&\Gamma^{(0)}_{2,2} = e_{0,1}e_{1,0}(
e_{0,1} e_{1,0} + e_{0,2}e_{2,0} + e_{1,0} e_{0,1} + \nonumber \\
&&e_{1,2} e_{2,1} + e_{2,0}e_{0,2} + e_{2,1}e_{1,2} + e_{2,3}e_{3,2})
\label{G22}
\end{eqnarray}
Let us remark that in this expansion in walks, the number of walks of the first trace
is less then the number of walks in the second trace; for the first trace there
is a single walk $\gamma_1^{(0)} = e_{0,1} e_{1,0}$, while expanding the
second trace the walks can visit the vertices in the first walk, so there
are more cases; in fact there are $7$ walks $\gamma_2^{(0)}$.

Similarly one can compute the set of paths $\Gamma^{(0)}_{k_1,k_2}$ for
$\texttt{tr} A^{k_1} \texttt{tr} A^{k_2}$.
One has
\begin{eqnarray}
&&\langle \texttt{tr} A^{k_1} \texttt{tr} A^{k_2} \rangle_c = \nonumber \\
&& \sum_{(\gamma_1,\gamma_2) \in M \Gamma^{(0)}_{k_1,k_2}}
\langle (\texttt{tr} A^{k_1})_{\gamma_1} (\texttt{tr} A^{k_2})_{\gamma_2} \rangle -
\langle (\texttt{tr} A^{k_1})_{\gamma_1} \rangle \langle  (\texttt{tr} A^{k_2})_{\gamma_2} \rangle 
\label{Gk1k2}
\end{eqnarray}
where the contribution due to $(\gamma_1,\gamma_2)$ vanish due to factorization, in the
case in which $\gamma_1$ and $\gamma_2$ do not have an edge in common.

From Eqs. (\ref{a22c}, \ref{G22}, \ref{Gk1k2}) one gets
\begin{equation}
\langle \texttt{tr} A^2 \texttt{tr} A^2 \rangle_c = 2 N_2 (v_4 - v_2^2)
\label{a2a2r2}
\end{equation}
which is the first line in Eq. (\ref{twopt}).

Alternatively one can compute separately
\begin{equation}
\langle \texttt{tr} A^2 \texttt{tr} A^2 \rangle = 
 2 N_2 v_4 + 4 N_3 v_2^2 + N_4 v_2^2
\end{equation}
so that
\begin{equation}
\langle \texttt{tr} A^2 \texttt{tr} A^2 \rangle_c = 2 N_2 v_4 + 4 N_3 v_2^2 + N_4 v_2^2 - (N_2 v_2)^2
\label{a2a2r1}
\end{equation}
obtaining the same result as in Eq. (\ref{a2a2r2}).)
The advantage of the first derivation is that it
is naturally expressed in terms linear in the falling factorials.

$r$-trace correlators for $r > 2$ can be similarly computed.

In Appendix E we list the first few one-, two and three-trace
connected correlators.
The algorithm described above has been used to compute iteratively in Python $\Gamma^{(0)}_{k_1,\cdots,k_r}$,
and hence the correlators.

\section{Appendix B: Leading order of the $r$-trace connected correlator contributions containing $v_{2j}$}

\subsection{Properties of $T$}
The generating function of closed walks on tree graphs, in which all edges
are run exactly twice, is
\begin{equation}
T(x) = \frac{1 - \sqrt{1 - 4 \, x^{2}} }{2 \, x^{2}} =
        \sum_{m\ge 0} C_m x^{2m} = 1 + x^2 T(x)^2
\label{wt}
\end{equation}
the power of $x$ indicates the number of steps in a walk;
$C_j$ is the $j$-th Catalan number.

The following identity holds \cite{lang}
\begin{equation}
T(x)^s = \sum_{j \ge 0} \frac{s}{2j + s} \binom{2j + s}{j} x^{2j}
\label{tm1}
\end{equation}

From the recursion relation for the Chebyshev polynomials of the II kind
\begin{equation}
        U_i(\frac{y}{2}) = y U_{i-1}(\frac{y}{2}) - U_{i-2}(\frac{y}{2})
\end{equation}
it follows that $R_j$, defined in Eq. (\ref{Aj}), satisfies the recursion relation
\begin{equation}
R_j(y) = (y^2 - 2) R_{j-1}(y) - R_{j-2}(y)
\label{pj}
\end{equation}

Let us prove Eq. (\ref{Rj}); it is easy to verify for $j=1,2$; the general case follows
by induction using Eq. (\ref{pj})

\subsection{Path enumeration}
Consider a $r$-trace connected correlator
$\langle \prod_{i=1}^r \frac{Tr B^{k_i}}{n}\rangle_c$
with $k = \sum_{i=1}^r k_i$.
and a graph with $V$ vertices, $E$ edges
and $L$ loops contributing to it. One has
\begin{equation}
V = E - L + 1
\label{vel}
\end{equation}
The dependence from $n$ of its contribution to this correlator is
$\frac{n^{-r}}{(n-1)^{\frac{k}{2}}}N_V$; the leading term is $n^{-l}$ with
\begin{equation}
-l = V - \frac{k}{2} - r
\label{lv}
\end{equation}
Let us prove that the highest moment is $v_{2(l-r+2)}$.
Let $n_h$ be the number of edges run $h$ times, hence contributing a factor $v_h$ to
the average; $h$ is even; one has
\begin{equation}
E = \sum_{h \ge 2} n_h
\label{en}
\end{equation}
and
\begin{equation}
k = \sum_{h \ge 2} h n_h
\label{khn}
\end{equation}
From Eqs. (\ref{vel},\ref{lv},\ref{en},\ref{khn})
\begin{equation}
        \sum_{h \ge 2} n_h (h/2 - 1) + L - 1 + r = l
\label{ek}
\end{equation}
For given $l$, the highest moment $v_h$ appears once, $n_h = 1$, all
the other momenta are $v_2$, so $h' = 2$ and $\sum_{h'} n_{h'} (h'/2 - 1) = 0$,
and $L = 0$, so that Eq. (\ref{ek}) gives
\begin{equation}
h = 2(l-r+2)
\label{hmax}
\end{equation}

The highest moment $v_h$ with $h > 2$ corresponds, according to Eq.(\ref{hmax}),
to $l = r - 2 + \frac{h}{2}$.

For $r=1$ the walk is a tree graph in which all the edges apart one are run twice, the
remaining edge is run $h$ times, $h$ even. 
\vskip 0.3cm
{\bf Theorem} 1.
The generating function counting the number of walks on rooted trees, in which each edge is run twice, apart from a special edge which is run $2m$ times, with $m \ge 1$, is
\begin{equation}
f_m(x) = \sum_{n \ge m} \binom{2n}{n - m} x^{2n}
\label{fh1}
\end{equation}
Let us consider first the case $m \ge 2$.
A walk can start with an edge run twice, or with the edge run $h=2m$ times.

Consider the case in which the first edge is run twice.
The special edge $e$ can be inserted at the end of any step of the walk on a Wigner tree.
Let us consider one of the Wigner walks (i.e. walks in which each edge is run exactly
twice) with length $2k$, called $\gamma$;
let $\gamma'$ be $\gamma$ 
in which it is marked the step at the end of which there is the vertex $v$, to which $e$
is to be added. Let us consider the class of tree walks starting on $v$ with
a step on $e$ and ending with a step on $e$ returning to $v$.
These walks contain in particular the steps $s_1,\cdots, s_h$ on $e$.
Let $g$ be the generating function counting the number of these walks;
these walks start with $s_1$ and end with $s_{2m}$;
between the steps $s_i$ and $s_{i+1}$, $i=1,\cdots,h-1$ there can be a Wigner walk,
so that $g = x^h T(x)^{h-1}$. The corresponding contribution to $f_m(x)$
due to $\gamma'$ is
$x^{2k} g$, the one of all $\gamma'$ is $2k x^{2k} g$. Since there are $C_k$
walks $\gamma$ on length $2k$ in a Wigner tree, one gets a contribution
$x \frac{d T(x)}{d x} x^{h} T(x)^{h-1}$.

If the walk starts with the special edge, at each step of the walk on
the special edge, apart
from the start, one can insert a Wigner walk, so that one gets
$x^h T(x)^h$.

Therefore the generating function of the walks on trees, in
which each edge is run twice, apart from a special edge which is run $2m$ times is
\begin{equation}
        f_m(x) = x^{2m} T(x)^{2m-1} (x \frac{d T(x)}{d x} + T(x)) =
        \frac{x}{2m} \frac{d}{d x}(x^{2m} T(x)^{2m})
        \label{fh}
\end{equation}
From Eqs. (\ref{fh}, \ref{tm1}) one gets Eq. (\ref{fh1}).

Let us consider now the case $m=1$.
The generating function for the walks in which each edge is run twice
and any edge can be selected as special edge is (see Eq.(\ref{wt}))
\begin{equation}
\frac{x}{2} \frac{d T(x)}{d x}
\end{equation}
which gives Eq. (\ref{fh}) for $m=1$.

\vskip 0.3cm

In \cite{kh14} the case $m=2$ of Theorem 1 has been proved in Lemma $6.2$,
the cases $m=1$ and $m=3$ have also been proved; the general case is assumed
to be true.

From this Theorem and Eq. (\ref{hmax}) one obtains Eq. (\ref{s.2}) for $j \ge 2$.

Let us turn to the highest moment terms for $r > 1$.
As we saw above, the highest moment terms are those in which there is a
single edge which is run more than twice, and there are no loops.

In Appendix A we have introduced the set of walks $\Gamma^{(0)}_{k_1,\cdots,k_r}$.
Let us define
\begin{equation}
	\Gamma^{(0,r)} = \bigcup_{k_1,\cdots,k_r} \Gamma^{(0)}_{k_1,\cdots,k_r}
\end{equation}

\vskip 0.3cm

{\bf Lemma 1} Consider the paths formed by $r > 1$ walks in
$\Gamma^{(0,r)}$, such that each walk is on a tree graph and
such that all the walks pass through a special edge, while all  the other
edges of the tree graphs are run only twice by the path.
The generating function counting the number of these paths, with given number
of times the special edge is run, is
\begin{equation}
        \Phi_r(x_1,\cdots,x_r, z) = 2^{r-1}\sum_{m_1 \ge 1}\cdots \sum_{m_r \ge 1}
        f_{m_1}(x_1)\cdots  f_{m_r}(x_r) z^{2m_1 + \cdots + 2m_r}
\label{le1}
\end{equation}
From Theorem $1$, the generating function counting the walks on tree graphs, 
and how many times the  special edge is run on a walk, is given by
\begin{equation}
\phi(x, z) = \sum_{m \ge 1} f_m(x) z^{2m}
\label{phixz}
\end{equation}
In the case $r > 1$ the special edge of the walk corresponding to a trace
can also be run only twice; it is special in the sense that it is the
edge in common between the walks corresponding to the traces.
The generating function counting the paths and how many times the special
edge is run on the $r$ walks of the $r$-point function is
\begin{equation}
\Phi_r(x_1,\cdots,x_r, z) = 2^{r-1} \prod_{i=1}^r \phi(x_i, z)
\label{phixz2}
\end{equation}
The special edges can be joined with two different orientations,
leading to a factor $2^{r-1}$.
Eq. (\ref{phixz2}) gives Eq. (\ref{le1}).

The sum $\gamma_{2k_1,\cdots,2k_r}$ of the leading highest moment terms
  in each term in the $1/n$-expansion of the  $r$-trace connected correlator
\begin{equation}
        n^{-r} \langle \prod_{i=1}^r \texttt{tr} B^{2 k_i} \rangle_c
\end{equation}
is obtained from
$[x_1^{2k_1}]\cdots [x_r^{2k_r}]\Phi_r(x_1,\cdots,x_r, z)$
by replacing $z^{2m}$ with $\frac{v_{2m}}{v_2^m}$
and by adding the powers of $n$ in agreement with Eq. (\ref{hmax})
\begin{equation}
	\gamma_{2k_1,\cdots,2k_r} =     n^{2-r} 2^{r-1} \Big(\prod_{i=1}^r 
	\sum_{j_i=1}^{k_i} \binom{2 k_i}{k_i - j_i} n^{-j_i}\Big)
	\frac{v_{2j+\cdots+2j_r}}{v_2^{j_1+\cdots+j_r}}
\label{Gr}
\end{equation}
% alternative form
%\begin{eqnarray}
%&&\gamma_{2k_1,\cdots,2k_r} = 
%n^{2-r} 2^{r-1} \sum_{j=3}^{k_1+\cdots+k_r} n^{-j} \frac{v_{2 j}}{v_2^j}
%	\nonumber \\
%&& \sum_{j_i=1}^{k_1}\cdots \sum_{j_r=1}^{k_r} \delta_{j_1+\cdots+j_r,j}
%	\binom{2 k_1}{k_1 - j_1} \cdots \binom{2 k_r}{k_r - j_r} 
%\label{Gr}
%\end{eqnarray}
From this equation, and the fact that if there are traces of an odd
power of matrices the contribution is subleading, follows Eq. (\ref{DBr}).

\section{Appendix C: $1/n$-order one-point Green function}
The $1$-point Green function is
\begin{equation}
G(y) = \langle \frac{1}{n} \texttt{tr} \big(\frac{1}{y - B}\big) \rangle
\label{G1}
\end{equation}

At leading order it is
\begin{equation}
G(y) = \frac{1}{y} T(\frac{1}{y})
\label{G0}
\end{equation}

In this appendix we reproduce the $1/n$ contributions $S_2, S_3, S_4$
to the correlation function given in \cite{em}, using the same kind of argument
present in the proof of the Theorem in Appendix B.
We use a similar notation an in \cite{em}, but with $x$
replaced by $x^2$.
While in the text we consider random matrices with zero diagonal elements,
here we take them to be random variables, with
\begin{equation}
\langle (A_{i,i})^2 \rangle = s^2
\label{aii}
\end{equation}

Let us consider the contribution due to graphs with a single loop.
The number of walks on a $p$-gon, in which each edge is run twice,
are $p+1$: $p$ of them move clockwise for $k$ steps, with $k=1,\cdots,p$;
then anticlockwise for $p$ steps, finally clockwise for $p-k$ steps;
the last walk goes clockwise for $2 p$ steps.
Let us consider the walks with one loop, which start on a tree with $k$ edges;
one can insert the loop in the walk on the tree at the end of each step;
there are $2k$ such insertions, so the generating function of these insertions
is $x T'$. Let the loop be a $p$-gon. There are $p+1$ walks on it, each walk
having 2p steps. At the end of each step one can insert a tree. Therefore
the generating function of the number of walks of this kind is
\begin{equation}
\sum_{p \ge 3} (p+1) x^{2p} x \frac{d T}{d x} T^{2p-1}
\label{lp1}
\end{equation}
Let us consider the walks starting with the loop. In this case one can add a tree
at the end of each step, so the generating function of the number of walks of this kind is

\begin{equation}
\sum_{p \ge 3} (p+1) x^{2p} T^{2p}
\label{lp2}
\end{equation}

The sum of the generating functions in Eq. (\ref{lp1}, \ref{lp2}) is
\begin{equation}
	S_4 = \sum_{p \ge 3} (p+1) x^{2p} T^{2p-1}(x \frac{d T}{d x} + T)
\end{equation}
in agreement with \cite{em}, with $r=1$.
The $x^{2k}$ coefficient of the Taylor expansion in $S_4$ is the coefficient
of $N_k v_2^k$ of $\langle \texttt{tr} A^{2k} \rangle$, see Eq. (\ref{onept}) for $k \le 7$.

\vskip 0.5cm
The contribution due to tree walks with an edge run $4$ times is $f_2$ in Eq. (\ref{fh}).
At the leading order in the $1/n$-expansion the corresponding contribution is
\begin{equation}
S_2 = \frac{v_4}{v_2^2} x^4 T^3 (x \frac{d T}{dx} + T)
\label{oneptv4}
\end{equation}
as obtained in \cite{em}.

\vskip 0.5cm

Consider next the contribution due to a self-loop.
A self-loop gives a factor $x^2 \frac{s^2}{\sigma^2}$, where $\sigma^2 = v_2$.
Proceeding as in the case of the contribution due to a loop,
if the walk starts with a tree, one gets 
$x \frac{d T}{d x}$ for the possible places of insertion of the self-loop;
the self-loop is run twice, so at the end of the first step of the self-loop
one can insert a tree, so one gets $x^2 \frac{s^2}{\sigma^2} T x \frac{d T}{d x}$.
If the walk starts with the self-loop, a tree can be inserted at the end of each
step of the self-loop, so one gets $x^2 \frac{s^2}{\sigma^2} T^2$.
Hence the contribution of the self-loop is the term $S_2$ in \cite{em}
\begin{equation}
	S_3 = x^2 \frac{s^2}{\sigma^2} T (x \frac{d T}{d x} + T) 
\end{equation}

Finally in \cite{em} is given the generating function for the $\frac{1}{n}$
contribution to $\langle \texttt{tr} A^{2k} \rangle$, due to the terms coming
from the expansion of the falling factorial $N_{k+1}$; since we didn't expand
the falling factorials in Eq. (\ref{onept}), we don't need to consider this term.

\section{Appendix D: Leading order of the two-point connected correlation function}

Let us compute the leading term of the $1/n$-expansion of
\begin{equation}
	{\cal C}_2(x_1, x_2) =
\sum_{k_1,k_2} x_1^{k_1} x_2^{k_2} \langle \texttt{tr} B^{k_1} \texttt{tr} B^{k_2}\rangle_c
\label{twoptc}
\end{equation}

The two-point connected correlation function is
\begin{equation}
G_c(y_1, y_2) = \frac{1}{n^2}\frac{1}{y_1y_2} {\cal C}_2(\frac{1}{y_1},\frac{1}{y_2})
\label{GC2}
\end{equation}

Let us consider the contributions with one loop. 
The loop is run once in the walk $\gamma_1$ corresponding to the first trace and once
in the walk $\gamma_2$ corresponding to the second trace.
The enumeration of walks $\gamma_1$ containing a $r$-sided loop proceed as in
the proof of the Theorem in Appendix B.
The first walk can start with a loop, or with a tree;
if it starts with a $r$-edged loop, at the end of each subsequent step along the loop
a tree can start, giving $x_1^r T(x_1)^r$; if it starts with a tree, the loop can occur
on any vertex of the tree apart from the first, so one gets
$x_1 \frac{d T}{d x_1} T(x_1)^{r-1} x_1^r$; together they give
\begin{equation}
\frac{x_1}{r} \frac{d}{d x_1} (x_1 T(x_1))^r
\end{equation}
similarly for the second
walk. The loop of the second walk can start at each of the vertices of the
loop of the first loop, and proceed in either direction; this leads to a factor
$2 r$ for an $r$-loop.

In the corresponding term of ${\cal C}_2(x_1, x_2)$ the coefficient of Eq. (\ref{twoptc})
in $x_1^{k_1} x_2^{k_2}$ is multiplied by $N_{(k_1+k_2)/2}(n-1)^{-(k_1+k_2)/2}$ 
(where $N_{(k_1+k_2)/2}$ comes from the fact that a graph with $(k_1+k_2)/2$ edges
and one loop has $(k_1+k_2)/2$ vertices; the factor $(n-1)^{-(k_1+k_2)/2}$
comes from the conversion from matrix $B$ to matrix $A$)) 
which is $1$ at leading order.
The contribution of these terms to ${\cal C}_2(x_1,x_2)$ to leading order is
\begin{equation}
\sum_{r \ge 3} 2r \frac{x_1}{r} \frac{d}{d x_1} (x_1 T(x_1))^r
\frac{x_2}{r} \frac{d}{d x_2} (x_2 T(x_2))^r
\label{c2loop}
\end{equation}

Let us now consider the contribution in which $\gamma_1$ and $\gamma_2$ are Wigner
walks, which have a single edge in common; according to the Theorem in Appendix B
there is a factor $f(x_i)$ counting
the number of walks $\gamma_i$ with one selected edge; the two selected edges
can be identified with two orientations, giving a factor $2$.
In considering the contribution due to the walks $\gamma_1$ e $\gamma_2$
to the two-point connected correlator in Eq. (\ref{Gk1k2})), one gets a factor
\begin{equation}
\langle A_{i,j}^2 A_{i,j}^2 \rangle - \langle A_{i,j}^2\rangle^2 = v_4 - v_2^2 \ ;
\nonumber
\end{equation}
it follows that
the contributions of walks with a single edge in common to ${\cal C}_2(x_1, x_2)$ are
\begin{equation}
       2 (v_4v_2^{-2} - 1) f_1(x_1) f_1(x_2) =
        \frac{x_1x_2(v_4 v_2^{-2} - 1)}{2}\frac{d T(x_1)}{d x_1} \frac{d T(x_2)}{d x_2}
\label{twoptse}
\end{equation}

From Eq, (\ref{c2loop}, \ref{twoptse}) at leading order one has
\begin{equation}
{\cal C}_2(x_1, x_2) = x_1 x_2 \frac{\partial}{\partial x_1} \frac{\partial}{\partial x_2}
\Big( \frac{(v_4 v_2^{-2} - 1)}{2} T(x_1) T(x_2) +
\sum_{r \ge 3} \frac{2}{r} (x_1 x_2 T(x_1) T(x_2))^r \Big)
\label{c2s}
\end{equation}

Let us rewrite this expression to compare it with the literature.

From Eqs. (\ref{wt}, \ref{c2s}) one gets
\begin{eqnarray}
{\cal C}_2(x_1, x_2) &=& x_1 x_2 \frac{\partial}{\partial x_1} \frac{\partial}{\partial x_2}
        \big[-2 \log(1-x_1 T(x_1) x_2 T(x_2)) + \nonumber \\
        &&\frac{1}{2}(v_4v_2^{-2} - 3) T(x_1) T(x_2) -2 x_1 T(x_1) x_2 T(x_2)\big]
\label{c2res}
\end{eqnarray}

Using $\frac{d G(y)^2}{d y} = - x^2 \frac{d T(x)}{d x}$, 
where $G(y)$ is the lowest-order one-point function Eq. (\ref{G0}) and 
$x = \frac{1}{y}$,
Eqs. (\ref{GC2}, \ref{c2res}) give
\begin{equation}
n^2 G_c(y_1, y_2) = \frac{\partial}{\partial y_1} \frac{\partial}{\partial y_2}
\Big(-2 \log(1 - G(y_1)G(y_2))) +
        \frac{1}{2}(v_4v_2^{-2} - 3) G(y_1)^2 G(y_2)^2 -2 G(y_1) G(y_2)\Big)
\label{gc2}
\end{equation}
which is the two-point correlation function in the Wigner matrix ensemble
considered in this paper.

Let us now consider Wigner symmetric matrices with diagonal elements, see
Eq. (\ref{aii}).

Diagonal elements give self-loops.
Given a Wigner walk on a graph with $k_1$ edges, one can add a self-loop at the
start of the walk and at the end of each step of the walk; same for the second
walk; the two walks have in common the self-loop, so one gets a factor
$\frac{s^2 x_1 x_2}{v_2}$;
therefore the contribution to ${\cal C}_2(x_1,x_2)$ is
\begin{eqnarray}
\frac{s^2}{v_2}\sum_{k_1,k_2} x_1^{2 k_1+1} x_2^{2 k_2+1} (2k_1 + 1) C_{k_1} (2k_2 + 1) C_{k_2} =
\frac{s^2}{v_2} x_1 x_2 \frac{\partial}{\partial x_1} \frac{\partial}{\partial x_2}
         \big(x_1 T(x_1) x_2 T(x_2)\big) \nonumber
\end{eqnarray}
Adding this term to Eq. (\ref{c2res}) one gets, for $s^2 = 2 v_2$,
\begin{eqnarray}
n^2 G_c(y_1, y_2) = -2 \frac{\partial}{\partial y_1} \frac{\partial}{\partial y_2}
        \log(1 - G(y_1)G(y_2)) + \frac{1}{2}(v_4v_2^{-2} - 3)\frac{\partial}{\partial y_1}  \frac{\partial}{\partial y_2}
        G(y_1)^2 G(y_2)^2
\nonumber \\
\label{gc2kkp1}
\end{eqnarray}
that is the result given in \cite{az}, with $\tau_4 = v_4v_2^{-2} - 3$.
Using the identities
\begin{eqnarray}
&&\frac{d G(z)}{d z} = -\frac{G(z)^2}{1 - G(z)^2} \nonumber \\
&& G(z_1)G(z_2)(z_1 - z_2) = -(G(z_1)-G(z_2))(1 - G(z_1)G(z_2))
\end{eqnarray}
one obtains equivalently the expression found in Eq. (I.15) in 
\cite{kkp2}, where $G(z) = - w r(z)$, $w = \sqrt{v_2}$ and $\sigma = v_4v_2^{-2} - 3$
\begin{eqnarray}
        n^2 G_c(y_1, y_2) &&= \frac{2}{(1 - G(z_1)^2) (1 - G(z_2)^2)}
\big(\frac{G(z_1) - G(z_2)}{z_1 - z_2}  \big)^2 + \nonumber \\
	&& 2(v_4v_2^{-2} - 3)\frac{G(z_1)^3 G(z_2)^3}{(1 - G(z_1)^2) (1 - G(z_2)^2) }
\label{gc2kkp}
\end{eqnarray}

\section{Appendix E: Exact connected correlators at low orders.}
\subsection{Single-trace averages.}
\small
\begin{eqnarray}
\langle \texttt{tr} A^2\rangle =&& N_2 v_2\nonumber \\
\langle \texttt{tr} A^4\rangle =&& 2N_3 v_2^2 + N_2 v_4 \nonumber \\
\langle \texttt{tr} A^6\rangle =&& 5 N_4 v_2^3 + N_3(6 v_4 v_2 + 4 v_2^3) + N_2 v_6 \nonumber \\
\langle \texttt{tr} A^8\rangle =&& 14 N_5 v_2^4 + N_4(28 v_4 v_2^2  + 37 v_2^4) +
        N_3 (8v_6 v_2 + 6v_4^2 + 28v_4 v_2^2) + N_2 v_8
\nonumber \\
\langle \texttt{tr} A^{10}\rangle =&&
42 N_6 v_2^5 + 4 N_5 (30 v_4 v_2^3 + 59 v_2^5) +
         5 N_4(9 v_6 v_2^2 + 13 v_4^2 v_2 + 77 v_4 v_2^3 + 29 v_2^5) + \nonumber \\
        &&      10 N_3(v_8 v_2 + 2 v_4 v_6 + 5 v_6 v_2^2 + 9 v_4^2 v_2) + N_2 v_{10}
   \nonumber \\
\langle \texttt{tr} A^{12}\rangle =
&& 132 N_7 v_2^6 + N_6(495 v_4 v_2^4 + 1289 v_2^6) +
2 N_5(110 v_6 v_2^3 + 231 v_4^2 v_2^2  + 1656 v_4 v_2^4 + 1203 v_2^6) +
\nonumber \\
&& N_4 (66 v_8 v_2^2 + 57 v_4^3 + 1902 v_4^2 v_2^2 + 252 v_4 v_6 v_2 + 2439 v_4 v_2^4 + 
774 v_6 v_2^3 + 340 v_2^6) +
\nonumber \\
&& 2 N_3(6 v_{10} v_2 + 67 v_4^3 + 204 v_4 v_6 v_2 + 10 v_6^2 + 15 v_4 v_8 + 
39 v_8 v_2^2) + N_2 v_{12} \nonumber \\
\langle \texttt{tr} A^{14}\rangle =&& 429 N_8 v_2^7 + N_7(2002 v_4 v_2^5 + 6476 v_2^7) +
7 N_6(143 v_6 v_2^4 + 390 v_4^2 v_2^3 + 3278 v_4 v_2^5 + 3479 v_2^7) +
\nonumber \\
&& 14 N_5(26 v_8 v_2^3 + 63 v_4^3 v_2 + 1607 v_4^2 v_2^3 + 143 v_4 v_6 v_2^2 + 
3606 v_4 v_2^5 + 
521 v_6 v_2^4 + 1342 v_2^7) +
\nonumber \\
&& 7 N_4(13 v_{10} v_2^2 + 727 v_4^3 v_2 + 54 v_4^2 v_6 + 2919 v_4^2 v_2^3 + 
1329 v_4 v_6 v_2^2 + 41 v_6^2 v_2 + 62 v_4 v_8 v_2 +
\nonumber \\
&& 1257 v_4 v_2^5 + 830 v_6 v_2^4 + 193 v_8 v_2^3) + \nonumber \\
&& 14 N_3(v_{12} v_2 + 83 v_4^2 v_6 + 3 v_{10} v_4 + 40 v_6^2 v_2 + 55 v_4 v_8 v_2 + 
	5 v_6 v_8 + 8 v_{10} v_2^2) + N_2 v_{14}  \qquad
\label{onept}
\end{eqnarray}

\normalsize
The odd one-point functions vanish.

The highest moments in the coefficients of the $\frac{1}{n}$-expansion of
$\langle \texttt{tr} A^{2k} \rangle$ are given by the Theorem in Appendix B and 
Eq. (\ref{hmax})
\begin{equation}
\sum_{m=2}^{k} \binom{2 k}{k - m} N_{k + 2 - m} v_{2m} v_2^{k - m}
\end{equation}

\subsection{Two-trace connected correlators}

Here are the results for the first few non-vanishing connected correlators

\small
\begin{eqnarray}
\langle \texttt{tr} A^2 \texttt{tr} A^2 \rangle_c =&& 2 N_2 (v_4 - v_2^2) \nonumber \\
\langle \texttt{tr} A^4 \texttt{tr} A^2 \rangle_c =&&
        8 N_3 (v_4 v_2- v_2^3) + 2 N_2 (v_6 - v_4 v_2)
        \nonumber \\
\langle \texttt{tr} A^3 \texttt{tr} A^3 \rangle_c =&& 6 N_3 v_2^3 \nonumber \\
\langle \texttt{tr} A^6 \texttt{tr} A^2 \rangle_c =&&
30 N_4 (v_4 v_2^2 - v_2^4) + 12 N_3(v_6 v_2 + v_4^2  - 2 v_2^4)  +
2 N_2 (v_8 - v_6 v_2) \nonumber \\
\langle \texttt{tr} A^5 \texttt{tr} A^3 \rangle_c =&&
30 N_4 v_2^4 + 30 N_3 v_4 v_2^2 \nonumber \\
\langle \texttt{tr} A^4 \texttt{tr} A^4 \rangle_c =&&
8N_4 (4 v_4 v_2^2 - 3 v_2^4) + 8 N_3(2 v_6 v_2 + v_4^2 - 3 v_2^4) +
2N_2 (v_8 - v_4^2)
\nonumber \\
\nonumber \\
\langle \texttt{tr} A^8 \texttt{tr} A^2 \rangle_c =&&
112 N_5(v_4 v_2^3 - v_2^5) +
8 N_4(7 v_6 v_2^2 + 14 v_4^2 v_2 + 16 v_4 v_2^3 - 37 v_2^5) + \nonumber \\
&& 8 N_3(2 v_8 v_2 + 3 v_6 v_2^2 + 5 v_4 v_6 + 11 v_4^2 v_2  - 21 v_4 v_2^3) +
2 N_2(v_{10} - v_8 v_2)
 \nonumber \\
\langle \texttt{tr} A^7 \texttt{tr} A^3 \rangle_c =&&
126 N_5 v_2^5 + 84 N_4 (3 v_4 v_2^3 + 2 v_2^5) + 42 N_3 (v_6 v_2^2 + 2 v_4^2 v_2)
 \nonumber \\
\langle \texttt{tr} A^6 \texttt{tr} A^4 \rangle_c =&&
24 N_5 (5v_4 v_2^3 - 3v_2^5) +
6 N_4(13 v_6 v_2^2 + 16 v_4^2 v_2 + 27 v_4 v_2^3 - 36 v_2^5) + \nonumber \\
&& 4 N_3(5v_8 v_2 + 9v_4v_6 + 10v_6 v_2^2 + 12v_4^2 v_2 - 24v_4 v_2^3   - 12 v_2^5) +
\nonumber \\
&& 2N_2 (v_{10} - v_4 v_6) \nonumber \\
\langle \texttt{tr} A^5 \texttt{tr} A^5 \rangle_c =&&
160 N_5 v_2^5 + 50 N_4 (7 v_4 v_2^3 + 2 v_2^5) + 50 N_3 (v_6 v_2^2 + 2 v_4^2 v_2)
 \nonumber \\
\langle \texttt{tr} A^{10} \texttt{tr} A^2 \rangle_c =&&
420 N_6 (v_4 v_2^4 - v_2^6) + 40 N_5 (6 v_6 v_2^3 + 18 v_4^2 v_2^2 + 35 v_4 v_2^4 - 
	59 v_2^6) + \nonumber \\
&& 10 N_4 (9 v_8 v_2^2 + 13 v_4^3 + 192 v_4^2 v_2^2 + 44 v_4 v_6 v_2 - 163 v_4 v_2^4 + 
	50 v_6 v_2^3 - 145 v_2^6) + \nonumber \\
	&& 20 N_3 (v_{10} v_2 + 9 v_4^3 - 27 v_4^2 v_2^2 + 24 v_4 v_6 v_2 + 
	2 v_6^2 + 3 v_4 v_8 - 15 v_6 v_2^3 + 3 v_8 v_2^2) +
\nonumber \\
&& 2 N_2 (v_{12} - v_{10} v_2)
\nonumber \\
\langle \texttt{tr} A^9 \texttt{tr} A^3 \rangle_c =&&
504 N_6 v_2^6 + 54 N_5 (28 v_4 v_2^4 + 37 v_2^6) +
18 N_4 (24 v_6 v_2^3 + 63 v_4^2 v_2^2 + 132 v_4 v_2^4 + 26 v_2^6) + \nonumber \\
&& 6 N_3 (9 v_8 v_2^2 + 22 v_4^3 + 54 v_4 v_6 v_2)
\nonumber \\
\langle \texttt{tr} A^8 \texttt{tr} A^4 \rangle_c =&&
224 N_6 (2 v_4 v_2^4 - v_2^6) + 
16 N_5 (21 v_6 v_2^3 + 42 v_4^2 v_2^2 + 109 v_4 v_2^4 - 84 v_2^6) + \nonumber \\
&& 8 N_4 (15 v_8 v_2^2 + 10 v_4^3 + 201 v_4^2 v_2^2 + 58 v_4 v_6 v_2 - 23 v_4 v_2^4 + 
89 v_6 v_2^3 - 184 v_2^6) + \nonumber \\
&& 8 N_3 (3 v_{10} v_2 + 11 v_4^3 - 30 v_4^2 v_2^2 + 48 v_4 v_6 v_2 + 5 v_6^2 + 
7 v_4 v_8 - 42 v_4 v_2^4 - 12 v_6 v_2^3 + 10 v_8 v_2^2) + \nonumber \\
&& 2 N_2 (v_{12} - v_4 v_8)
\nonumber \\
\langle \texttt{tr} A^7 \texttt{tr} A^5 \rangle_c =&&
700 N_6 v_2^6 + 70 N_5 (33 v_4 v_2^4 + 28 v_2^6) +
70 N_4 (9 v_6 v_2^3 + 24 v_4^2 v_2^2 + 32 v_4 v_2^4 + 4 v_2^6) + \nonumber \\
&& 70 N_3 (v_8 v_2^2 + 2 v_4^3 + 6 v_4 v_6 v_2)
\nonumber \\
\langle \texttt{tr} A^6 \texttt{tr} A^6 \rangle_c =&&
150 N_6 (3 v_4 v_2^4 - v_2^6) + 
72 N_5 (5 v_6 v_2^3 + 9 v_4^2 v_2^2 + 26 v_4 v_2^4 - 15 v_2^6) + \nonumber \\
&& 6 N_4 (22 v_8 v_2^2 + 19 v_4^3 + 243 v_4^2 v_2^2 + 72 v_4 v_6 v_2 + 57 v_4 v_2^4 + 
122 v_6 v_2^3 - 235 v_2^6) +
\nonumber \\
&& 12 N_3 (2 v_{10} v_2 + 11 v_4^3 - 18 v_4^2 v_2^2 + 26 v_4 v_6 v_2 + 3 v_6^2 + 
5 v_4 v_8 - 24 v_4 v_2^4 - 4 v_6 v_2^3 + 7 v_8 v_2^2 - 8 v_2^6) +
\nonumber \\
&& 2 N_2 (v_{12} - v_6^2)
\label{twopt}
\end{eqnarray}

\normalsize
\vskip 0.6cm

From Eqs. (\ref{tm1}),(\ref{fh1}),(\ref{c2s}) one has that the leading term $(n v_2)^{\frac{m_1+m_2}{2}}$ 
in  $\langle \texttt{tr} A^{m_1} \texttt{tr} A^{m_2}\rangle_c$, with $m_1+m_2$ even, has coefficient
\begin{equation}
- 2 \binom{m_1}{\frac{m_1}{2}-1}\binom{m_2}{\frac{m_2}{2}-1}\delta_{m_1,even} +
\sum_{r \ge 3; m_1-r\, even} 2r \binom{m_1}{\frac{m_1-r}{2}} \binom{m_2}{\frac{m_2-r}{2}}
\end{equation}

The highest moments $v_{2m}$, where $m \ge 2$, in the coefficients of the $\frac{1}{n}$-expansion of
$\langle \texttt{tr} A^{2 k_1} \texttt{tr} A^{2 k_2} \rangle_c$ are given by Eq. (\ref{Gr}), i.e. by
\begin{equation}
\sum_{m=2}^{k_1 + k_2} n^{k_1+k_2+2-m} v_{2m} v_2^{k_1 + k_2 - m}
\, 2 \sum_{m_1=1}^{k_1}  \sum_{m_2=1}^{k_2} \delta_{m_1 + m_2, m}
\binom{2k_1}{k_1-m_1} \binom{2k_2}{k_2-m_2}
\label{twopt1}
\end{equation}

$\langle \texttt{tr} A^{m_1} \texttt{tr} A^{m_2} \rangle_c$ with $m_1, m_2$ odd have
all highest moments $v_{2m}$, where $m \ge 2$, with coefficient having a power of 
$n$ lower than $\frac{m_1+m_2}{2}+2-m$.

\subsection{Three-point connected correlation functions}
The first few non-vanishing connected correlators are

\small
\begin{eqnarray}
\langle \texttt{tr} A^2 \texttt{tr} A^2 \texttt{tr} A^2 \rangle_c =&& 4 N_2(v_6 - 3 v_4 v_2 + 2 v_2^3)
        \nonumber \\
\langle \texttt{tr} A^4 \texttt{tr} A^2 \texttt{tr} A^2 \rangle_c =&&
        16 N_3(v_6 v_2 + v_4^2 - 5v_4 v_2^2 + 3 v_2^4) +
        4 N_2(v_8 - 2v_6 v_2 - v_4^2 + 2v_4 v_2^2)
\nonumber \\
\langle \texttt{tr} A^3 \texttt{tr} A^3 \texttt{tr} A^2 \rangle_c =&& 36 N_3(v_4 v_2^2 - v_2^4)
\nonumber \\
\langle \texttt{tr} A^6 \texttt{tr} A^2 \texttt{tr} A^2 \rangle_c =&&
60 N_4(v_6 v_2^2 + 2 v_4^2 v_2 - 7v_4 v_2^3 + 4 v_2^5) + \nonumber \\
&& 24 N_3(v_8 v_2 + 3v_4 v_6 - 2 v_6 v_2^2 - 2v_4^2 v_2 - 8 v_4 v_2^3 + 8 v_2^5) +
\nonumber \\
&& 4 N_2(v_{10} - 2v_8v_2 - v_6 v_4 + 2v_6 v_2^2) \nonumber \\
\langle \texttt{tr} A^5 \texttt{tr} A^3 \texttt{tr} A^2 \rangle_c =&&
240 N_4 (v_4 v_2^3 - v_2^5) + 60 N_3 (v_6 v_2^2 + 2 v_4^2 v_2 - 3 v_4 v_2^3)
\nonumber \\
\langle \texttt{tr} A^4 \texttt{tr} A^4 \texttt{tr} A^2 \rangle_c =&&
64 N_4(v_6 v_2^2 + 2 v_4^2 v_2 - 6 v_4 v_2^3 + 3 v_2^5) + \nonumber \\
&& 32 N_3(v_8 v_2 + 2 v_4 v_6 - 2 v_6 v_2^2 - v_4^2 v_2 - 6 v_4 v_2^3 + 6v_2^5) +
\nonumber \\
&& 4 N_2( v_{10} - v_8 v_2 - 2 v_4 v_6 + 2 v_4^2 v_2) \nonumber \\
\langle \texttt{tr} A^4 \texttt{tr} A^3 \texttt{tr} A^3 \rangle_c =&&
144 N_4 v_4 v_2^3 + 36 N_3 (v_6 v_2^2 + 2 v_4^2 v_2 - v_4 v_2^3 - 2 v_2^5)\nonumber \\
\langle \texttt{tr} A^8 \texttt{tr} A^2 \texttt{tr} A^2 \rangle_c =&&
224 N_5(v_6 v_2^3 + 3 v_4^2 v_2^2 - 9 v_4 v_2^4 + 5 v_2^6) + \nonumber \\
&& 16 N_4(7 v_8 v_2^2 + 14 v_4^3 + 6 v_4^2 v_2^2 + 42 v_4 v_6 v_2 - 249 v_4 v_2^4
- 5 v_6 v_2^3 +  185 v_2^6) +
\nonumber \\
&& 16 N_3(2 v_{10} v_2 + 11 v_4^3 - 96 v_4^2 v_2^2 + 18 v_4 v_6 v_2 + 5 v_6^2 + 
7 v_4 v_8 + 84 v_4 v_2^4 - 30 v_6 v_2^3 - v_8 v_2^2) + \nonumber \\
&& 4 N_2(v_{12} - v_4 v_8 - 2 v_{10} v_2  + 2 v_8 v_2^2) \nonumber \\
\langle \texttt{tr} A^7 \texttt{tr} A^3 \texttt{tr} A^2 \rangle_c =&&
1260 N_5 (v_4 v_2^4 - v_2^6) + 
168 N_4 (3 v_6 v_2^3 + 9 v_4^2 v_2^2 - 2 v_4 v_2^4 - 10 v_2^6) + \nonumber \\
&& 84 N_3(v_8 v_2^2 + 2 v_4^3 - 6 v_4^2 v_2^2 + 6 v_4 v_6 v_2 - 3 v_6 v_2^3) \nonumber \\
\langle \texttt{tr} A^6 \texttt{tr} A^4 \texttt{tr} A^2 \rangle_c =&&
240 N_5 (v_6 v_2^3 + 3 v_4^2 v_2^2 - 7 v_4 v_2^4 + 3 v_2^6) + \nonumber \\
&& 12 N_4 (13 v_8 v_2^2 + 16 v_4^3 + 33 v_4^2 v_2^2 + 58 v_4 v_6 v_2 - 288 v_4 v_2^4 - 
	12 v_6 v_2^3 + 180 v_2^6)  + \nonumber \\
&& 8 N_3(5 v_{10} v_2 + 12 v_4^3 - 108 v_4^2 v_2^2 + 26 v_4 v_6 v_2 + 9 v_6^2 + 
14 v_4 v_8 + 36 v_4 v_2^4 - 54 v_6 v_2^3 + 60 v_2^6) + \nonumber \\
&& 4 N_2 (v_{12} + 2 v_4 v_6 v_2 - v_6^2 - v_4 v_8 - v_{10} v_2) \nonumber \\
\langle \texttt{tr} A^6 \texttt{tr} A^3 \texttt{tr} A^3 \rangle_c =&&
540 N_5(v_4 v_2^4 + v_2^6) + 216 N_4(v_6 v_2^3 + 3 v_4^2 v_2^2 + 8 v_4 v_2^4 - 2 v_2^6) +
\nonumber \\
&& 36 N_3 (v_8 v_2^2 + 4 v_4^3 + 6 v_4 v_6 v_2 - 6 v_4 v_2^4 - v_6 v_2^3 - 4 v_2^6) 
\nonumber \\
\langle \texttt{tr} A^5 \texttt{tr} A^5 \texttt{tr} A^2 \rangle_c =&&
1600 N_5(v_4 v_2^4 - v_2^6) + 
100 N_4 (7 v_6 v_2^3 + 21 v_4^2 v_2^2 - 18 v_4 v_2^4 - 10 v_2^6) + \nonumber \\
&& 100 N_3(v_8 v_2^2 + 2 v_4^3 - 6 v_4^2 v_2^2 + 6 v_4 v_6 v_2 - 3 v_6 v_2^3)
\nonumber \\
\langle \texttt{tr} A^5 \texttt{tr} A^4 \texttt{tr} A^3 \rangle_c =&&
960 N_5 v_4 v_2^4 + 120 N_4 (4 v_6 v_2^3 + 9 v_4^2 v_2^2 + 8 v_4 v_2^4 - 9 v_2^6) + 
\nonumber \\
&& 60 N_3 (v_8 v_2^2 + 2 v_4^3 - 3 v_4^2 v_2^2 + 6 v_4 v_6 v_2 - 6 v_4 v_2^4)
\nonumber \\
\langle \texttt{tr} A^4 \texttt{tr} A^4 \texttt{tr} A^4 \rangle_c =&&
256 N_5 (v_6 v_2^3 + 3 v_4^2 v_2^2 - 6 v_4 v_2^4 + 3 v_2^6) +
64 N_4 (3 v_8 v_2^2 + v_4^3 + 12 v_4^2 v_2^2 + 12 v_4 v_6 v_2 - \nonumber \\
&& 54 v_4 v_2^4 - 
3 v_6 v_2^3 + 31 v_2^6) + \nonumber \\
&& 16 N_3 (3 v_{10} v_2 -2 v_4^3 - 27 v_4^2 v_2^2 + 12 v_4 v_6 v_2 + 5 v_6^2 + 
6 v_4 v_8 - 36 v_6 v_2^3 + 3 v_8 v_2^2 + 36 v_2^6)
+ \nonumber \\
&& 4 N_2 (v_{12} + 2 v_4^3 - 3 v_4 v_8)
\label{threept}
\end{eqnarray}

\normalsize
For $m \ge 3$ the highest moments $v_{2m}$ in the coefficients of the 
$\frac{1}{n}$-expansion of
$\langle \texttt{tr} A^{2 k_1} \texttt{tr} A^{2 k_2} \texttt{tr} A^{2 k_3}\rangle_c$ are given by Eq. (\ref{Gr}), i.e. by
\begin{eqnarray}
&&\sum_{m=3}^{k_1 + k_2 + k_3} n^{k_1+k_2+k_3+2-m} v_{2m} v_2^{k_1 + k_2 + k_3- m}
\label{threept1} \\
&& 4 \sum_{m_1=1}^{k_1}  \sum_{m_2=1}^{k_2} \sum_{m_3=1}^{k_3} \delta_{m_1 + m_2 + m_3, m}
\binom{2k_1}{k_1-m_1} \binom{2k_2}{k_2-m_2} \binom{2k_3}{k_3-m_3}
\nonumber
\end{eqnarray}

$\langle \texttt{tr} A^{m_1} \texttt{tr} A^{m_2} \texttt{tr} A^{m_3} \rangle_c$ with $m_1, m_2$ and $m_3$ not all
even have
all highest moments $v_{2m}$, where $m \ge 2$, with coefficient having a power of 
$n$ lower than $\frac{m_1+m_2+m_3}{2}+2-m$.


\vskip 2cm

\begin{thebibliography}{99} \bibitem {ge} G.'t Hooft,  \textsl{A planar
diagram theory for  strong interactions}, Nucl. Phys. \textbf{B72} 461
(1974).

\bibitem{bre1}E. Brezin, C. Itzykson, G. Parisi, J. B. Zuber
\textsl{Planar diagrams}, Commun.Math. Phys. \textbf{59}, 35–51
(1978). https://doi.org/10.1007/BF01614153.

\bibitem{amb} J. Ambjorn, B. Durhuus, T. Jonsson, \textsl{Quantum
Geometry, A statistical field theory approach}, Cambridge Univ. Press,
1997.

\bibitem{diF} P.Di Francesco, P. Ginsparg, J. Zinn Justin. \textsl{2 D
gravity and random matrices}, Physics Reports \textbf{254} (1995), 1-133.

\bibitem{bre2} E. Brezin and A. Zee, \textsl{Universality of the
correlations between eigenvalues of large random matrices},
Nucl. Phys. \textbf{B402} (1993) 613-627. E. Brezin and
A. Zee, \textsl{Correlation functions in disordered systems},
Phys. Rev. \textbf{E49}, (1994) 2588. E. Brezin, S. Hikami and
A. Zee, \textsl{Universal correlations for deterministic plus random
Hamiltonians}, Phys. Rev. \textbf{E51}, (1995) 5442  E. Brezin and A. Zee,
\textsl{Universal relations between Green functions in random matrix
theory}, Nucl. Phys. \textbf{B453} (1995) 531-551. G. Hackenbroich and
H. A. Weidenm\"uller, \textsl{Universality of Random-Matrix Results for
Non-Gaussian Ensembles}, Phys. Rev. Lett. \textbf{74}, (1995) 4118.

\bibitem{yau}  Horng-Tzer Yau, \textsl{The Wigner-Dyson-Gaudin-Mehta
Conjecture}, Notices of ICCM \textbf{1} (2013) 10-13.

\bibitem{dei} P. Deift, \textsl{Orthogonal Polynomials and Random
Matrices: a Riemann-Hilbert Approach}, Amer. Math. Soc., Providence,
R. I. 1999. P. Deift, D. Gioev, \textsl{Random Matrix Theory: Invariant
Ensembles and Universality}, Courant Lecture Notes 18, Amer. Mathematical
Soc. (2009).

\bibitem{ben} G. Ben Arous and A. Guionnet, \textsl{Wigner  matrices},
in  The Oxford Handbook of Random Matrix Theory, Eds. G. Akemann, J. Baik,
and P. Di Francesco (Oxford University Press, Oxford, 2011).

\bibitem{az} J. D'Anna and A. Zee, \textsl{Correlations  between
eigenvalues of large random matrices with independent entries},
Phys. Rev. \textbf{E 53}, 1399-1410 (1996).

\bibitem{kkp1} A. Khorunzy, B. Khoruzhenko and L. Pastur, \textsl{On
the $1/N$ corrections to the Green functions of random matrices with
independent entries}, J. Phys. \textbf{A28} (1995) L31.

\bibitem{kkp2} A. Khorunzy, B. Khoruzhenko and L. Pastur, \textsl{
Asymptotic properties of large random matrices with independent entries
  }, J. Math. Phys. \textbf{37} (1996) 5033.

\bibitem{tao} T. Tao, Van Vu, \textsl{Random matrices: Universality
of local eigenvalue statistics}, Acta Mathematica \textbf{206} (2011)
127-204.

\bibitem{flo} F. Benaych-Georges, A. Knowles, \textsl{Lectures on the
local semicircle law for Wigner matrices}, arXiv:1601.04055v4 [math.PR].

\bibitem{alice} A. Guionnet , \textsl{Large Random Matrices: Lectures on Macroscopic Asymptotics}, 
 Ecole d’ \'Et\'e de Probabilit\'es
 de Saint-Flour XXXVI–2006, Springer 2008.

\bibitem{ci} G. M. Cicuta \textsl{Real symmetric random matrices and
path counting}, Phys. Rev.\textbf{E72}, 026122 (2005).

\bibitem{tao2} T. Tao,  \textsl{Topics in Random Matrix}, Graduate Studies in Mathematics, Amer. Math. Soc. (2010).

\bibitem{for}  P. J. Forrester and A.K. Trinh, \textsl{Comment on "`Finite
size effects in the averaged eigenvalue density of Wigner random-sign
real symmetric matrices"'}, Phys. Rev. \textbf{E99}, 036101 (2019).

\bibitem{em}  N. Enriquez and L. Menard, \textsl{Asymptotic expansion of the expected
spectral measure of Wigner matrices}, Electron. Commun. Probab. {\bf 21}. 1 (2016).

\bibitem{lang} W. Lang, \textsl{On polynomials related to powers of the generating function of Catalan's
	numbers}, Fibonacci Quarterly \textbf{38}, 408 (2000).

\bibitem{bau} M. Bauer and O. Golinelli, \textsl{Random incidence matrices: moments of the spectral density},J. Stat. Phys. \textbf{103}, 301 (2001).

\bibitem{kh14}  A. Khorunzy, \textsl{On High Moments and the Spectral Norm of Large
Dilute Wigner Random Matrices }, journ. Math. Phys. {\bf 10}, 64 (2014).

\end{thebibliography}
\end{document}